\documentclass[aps,prd,groupedaddress,showpacs,tighten,floats,twocolumn,nofootinbib]{revtex4}
\usepackage{graphicx}
\usepackage{latexsym}
\def\beq{\begin{equation}}
\def\eeq{\end{equation}}
\def\bey{\begin{eqnarray}}
\def\eey{\end{eqnarray}}

\def\lsim{\mathrel{\raise.3ex\hbox{$<$\kern-.75em\lower1ex\hbox{$\sim$}}}}
\def\gsim{\mathrel{\raise.3ex\hbox{$>$\kern-.75em\lower1ex\hbox{$\sim$}}}}

\begin{document}

\title{Detecting MeV Gauge Bosons With High-Energy Neutrino Telescopes}  
\author{Dan Hooper}
\address{Fermi National Accelerator Laboratory, Theoretical Astrophysics, Batavia, IL  60510}

\date{\today}

\begin{abstract}

If annihilating MeV-scale dark matter particles are responsible for the observed 511 keV emission from the Galactic bulge, then new light gauge bosons which mediate the dark matter annihilations may have other observable consequences. In particular, if such a gauge boson exists and has even very small couplings to Standard Model neutrinos, cosmic neutrinos with $\sim$TeV energies will scatter with the cosmic neutrino background through resonant exchange, resulting in a distinctive spectral absorption line in the high-energy neutrino spectrum. Such a feature could potentially be detected by future high-energy neutrino telescopes.

\end{abstract}
\pacs{95.85.Ry;95.30.Cq,95.35.+d; FERMILAB-PUB-07-017-A}
\maketitle

\section{Introduction}

Some years ago, it was suggested that the 511 keV radiation observed from the Galactic bulge by the SPI/INTEGRAL satellite \cite{integral} may be the product of dark matter annihilations \cite{511dark}. For dark matter particles to generate the observed spectral line width of this signal, their annihilations must inject positrons with energies below a few MeV \cite{beacom} and, therefore, such particles must be much lighter than the range of masses typically associated with dark matter.

It has long been understood that weakly interacting particles with masses smaller than a few GeV (but larger than $\sim$1 MeV) are expected to be overproduced in the early universe relative to the measured dark matter abundance \cite{lee}. This conclusion can be modified, however, if new light mediators are introduced which make dark matter annihilations more efficient \cite{lightok}. For example, neutralinos within the MSSM are required by this argument to be heavier than $\sim$20 GeV \cite{susycase}. Within extended supersymmetric models with light Higgs bosons mediating neutralino annihilation, however, much smaller masses are possible~\cite{nmssm}.

For dark matter particles with 0.5-3 MeV masses to generate the measured dark matter abundance, they must annihilate during the freeze-out epoch with a cross section of $\sigma v \sim$ pb. To inject the flux of positrons needed to generate the signal observed by SPI/INTEGRAL, however, an annihilation cross section of $\sigma v \sim 10^{-4}-10^{-5}$ pb is required. Together, these requirements lead us to consider dark matter particles with s-wave suppressed annihilations ($\sigma v \propto v^2$). Such behavior can be found, for example, in the case of fermionic or scalar dark matter particles annihilating through a vector mediator. For such a dark matter particle to annihilate sufficiently in the early universe to not be overabundant today, the mediating boson would also have to be quite light~\cite{scalar,fayet}.

The existence of a new gauge boson which couples to dark matter and (at least some of the) Standard Model fermions would have a number of potentially observable consequences. Constraints on such a scenario have been placed by collider experiments \cite{collider}, neutrino experiments \cite{boehmneutrino,scalar}, atomic physics experiments \cite{atomic} and by observations of core-collapse supernovae \cite{sn}. 

In this letter, we present a new way in which a gauge boson associated with MeV-scale dark matter may be detected. In particular, neutrinos produced in cosmic ray accelerators could interact with cosmic background neutrinos via the resonant exchange of the new gauge boson, resulting in an absorption line in the high-energy cosmic neutrino spectrum. This resembles, in some respects, the scattering of ultra-high energy neutrinos through resonant $Z$-exchange, known as $Z$-burst mechanism~\cite{zburst}. Similar ideas have also been explored for very light gauge bosons associated with the generation of neutrino masses \cite{goldberg}. We find that if the new gauge boson's coupling to neutrinos is larger than approximately $g_{U \nu \nu} \gsim 10^{-5} (m_U/\rm{MeV})$, then this process will efficiently deplete the neutrino flux over a range of energies, $m^2_U/2m_{\nu}(1+z) < E_{\nu} < m^2_U/2 m_{\nu}$, where $z$ is the redshift of the neutrino source and $m_U$ is the mass of the gauge boson. Although challenging, such a feature could potentially be observed by future high-energy neutrino telescopes.

\section{Neutrino Absorption Via A Light Gauge Boson Resonance}

If a new gauge boson exists which couples to neutrinos, then propagating neutrinos will scatter with the cosmic neutrino background at a resonant energy of $E_{\nu} \approx m^2_U/2 m_{\nu}$, where $m_U$ is the mass of the new gauge boson. For a gauge boson with an MeV-scale mass, and considering $\sim$0.1 eV neutrinos, this corresponds to a resonance at $E_{\nu}\sim$1-10 TeV.

Although the couplings of a light gauge boson to neutrinos are quite tightly constrained by $\nu e$ scattering experiments~\cite{scalar,fayet}, over very long baselines, high-energy neutrinos may interact efficiently with cosmic background neutrinos via the exchange of the new boson, even if the relevant couplings are very small. These resonant interactions will be elastic in nature ($\nu \bar{\nu} \rightarrow \nu \bar{\nu}$) or, if $m_U > 2 m_e$, produce electron-positron pairs ($\nu \bar{\nu} \rightarrow e^+ e^-$). We first focus on the case in which $m_U < 2 m_e$. 

$\nu \bar{\nu} \rightarrow \nu \bar{\nu}$ processes will deplete the number of neutrinos in the resonance range, leading to an absorption line and a slight ``pile up" feature in the cosmic neutrino spectrum. Near resonance, the neutrino-antineutrino cross section is well approximated by
\begin{equation}
\sigma_{\nu \bar{\nu}} \approx  \frac{g^4_{U \nu \nu} s}{16 \pi [(m^2_U-s)^2+m^2_U \Gamma^2_U]},
\end{equation}
where $g_{U \nu \nu}$ is the $U$'s coupling to neutrinos. The decay width of the gauge boson is given by:
\begin{equation}
\Gamma_U \approx \sum_{\nu} \frac{g^2_{U \nu \nu} m_U}{12 \pi},
\end{equation}
where the sum is over the three neutrino species. 
At the resonant energy, this leads to a cross section of 
\begin{equation}
\sigma_{\nu \bar{\nu}} \sim \frac{\pi}{m^2_U},
\end{equation}
and a corresponding mean free path for a neutrino of
\begin{equation}
\lambda \approx \frac{1}{n_{\nu} \sigma_{\nu \bar{\nu}}} \sim \rm{pc} \, \bigg(\frac{m_U}{1\, \rm{MeV}}\bigg)^2,
\end{equation}
where $n_{\nu}= 3 \pi \Gamma(3) \zeta(3) T^3_{\nu} \approx 56 (1+z)^3 \rm{cm}^{-3}$ is the number density of cosmic background neutrinos (per flavor). 

%

Although this mean free path is very short by the standards of neutrino astronomy, the width of the corresponding resonance is also very narrow. For a 1 MeV gauge boson and a 0.1 eV neutrino mass, a neutrino will scatter on resonance only within a small range of energies: $\Delta E_{\nu} \sim 1 \, \rm{TeV} \times g^2_{U \nu \nu}$. The neutrino's energy moves relative to the resonance as a result of redshift energy losses, however, effectively broadening the resonance considerably \cite{goldberg}. The depth of such a resonance will be sufficient to deplete the neutrino flux over an energy range $m^2_U/(2 m_{\nu} (1+z)) < E_{\nu} < m^2_U/(2 m_{\nu})$ if the distance propagated while on resonance is larger than the mean free path:
\begin{eqnarray}
\Delta D \gsim \lambda \Longrightarrow \frac{c}{H}\frac{\Delta E_{\nu}}{E_{\nu}} \approx \frac{g^2_{U \nu \nu} c}{4 \pi H} \gsim \frac{m^2_U}{\pi n_{\nu}},
\end{eqnarray}
where $H$ is the rate of Hubble expansion. This condition is satisfied if the $U-\nu-\nu$ coupling is larger than:
\begin{equation}
g_{U \nu \nu} \gsim 4 \times 10^{-5} \bigg(\frac{m_U}{1\,\rm{MeV}}\bigg) \frac{1}{(1+z)^{3/2}},
\label{condition}
\end{equation}
where $z$ is the redshift at which the neutrino possesses the energy needed to scatter at resonance.

Even if the gauge boson is heavy enough to decay to $e^+ e^-$, the result of Eq.~\ref{condition} is expected to remain largely unchanged. Although the resonant cross section would be modified by a factor of $\sim (g_{U\nu \nu}/g_{Uee})^2$ by allowing such decays, the change in the width of the resonance will counteract this effect. Considering the $m_U > 2 m_e$ case, therefore, makes the condition of Eq.~\ref{condition} only slightly less restrictive.

\begin{figure}

\resizebox{8.5cm}{!}{\includegraphics{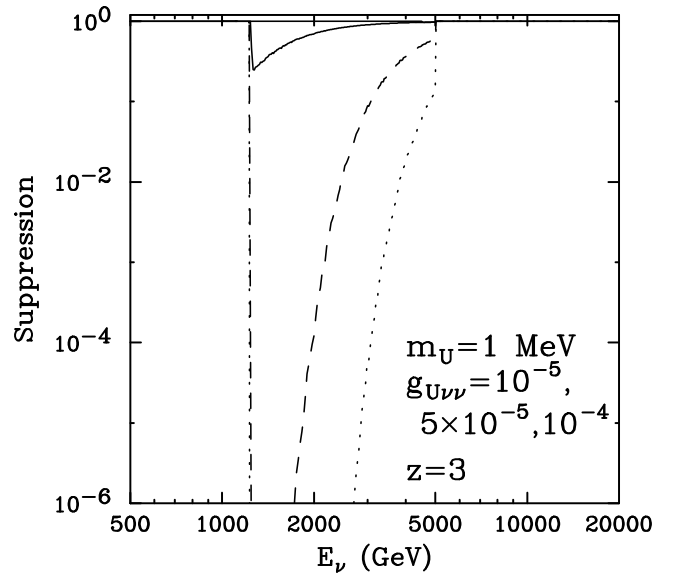}}
\caption{The suppression of the high-energy neutrino spectrum from a source at redshift $z=3$ for the case of $m_U=1$ MeV, and for three choices of the gauge boson's coupling to neutrinos: $g_{U\nu\nu}=10^{-5}$ (solid), $5 \times 10^{-5}$ (dashed) and $10^{-4}$ (dotted).}
\label{g}
\end{figure}

\begin{figure}

\resizebox{8.5cm}{!}{\includegraphics{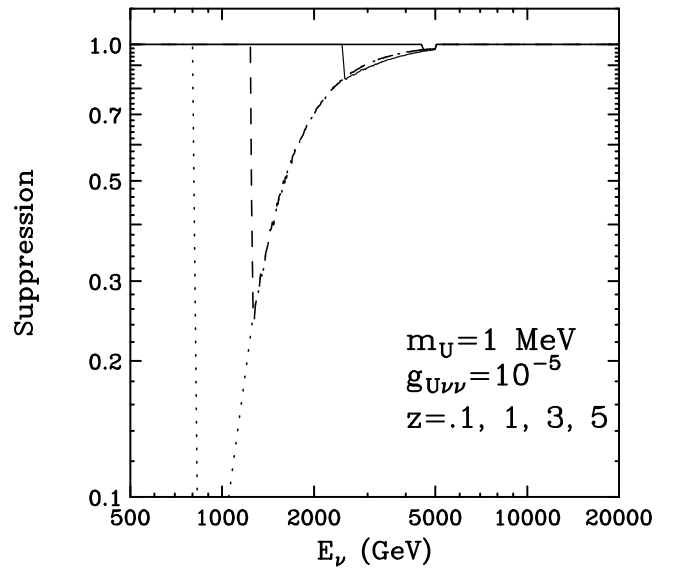}}
\caption{The suppression of the high-energy neutrino spectrum for the case of $m_U=1$ MeV, $g_{U\nu\nu}=10^{-5}$ and for sources at a redshift of $z=0.1$ (thick solid), 1.0 (thin solid), 3.0 (dashed) and 5.0 (dotted).}
\label{z}
\end{figure}

In Figs.~\ref{g} and \ref{z}, these results are illustrated. For a gauge boson with a $\sim\,$1 MeV mass and couplings to neutrinos of order $\sim$$10^{-5}$, high-redshift sources will have their neutrino flux considerably suppressed within the range $E_R/(1+z) < E_{\nu} < E_R$, where $E_R=m^2_U/2 m_{\nu}$ is the resonant energy.

\section{Discussion}

In the basic scenario we are considering here, there are a number of potentially free parameters, including the gauge boson's couplings to various fermions, the gauge boson's coupling to dark matter, and the masses of the gauge boson and dark matter particle. These quantities can be constrained by a number of considerations. 

Firstly, the product of the gauge boson's couplings to neutrinos and electrons is constrained by $\nu e$ scattering experiments, such that $g_{U \nu \nu} \sqrt{g^2_{U e_R e_R}+g^2_{U e_L e_L}} \lsim m^2_U G_F$ \cite{scalar,fayet}. To satisfy this constraint and at the same time generate the observed abundance of dark matter, some care must be taken. The most simple solution would be to set the $U$'s couplings to all left-handed fermions (including neutrinos) to zero. In this case, there will be no observable consequences in $\nu e$ experiments, or for  neutrino astronomy. On the other hand, if the $U$'s couplings to left and right-handed fermions are similar in magnitude (or the right-handed couplings are sub-dominant or zero) we find that these couplings can be as large as $g_{U \nu \nu} \sim g_{U e_L e_L} \lsim 4 \times 10^{-6} (m_U/\rm{MeV})$. If we consider the case of $g_{U \nu \nu}=g_{U e_L e_L}$, we can simultaneously satisfy $\nu e$ scattering and generate a significant degree of neutrino absorption ({\it ie.} satisfy the condition of Eq.~\ref{condition}) for high-redshift sources ($z \gsim 3$). If the coupling $g_{U \nu \nu}$ is somewhat larger than the corresponding coupling to electrons, less distant sources could also possess this feature.

\section{Prospects And Challenges For Future High-Energy Neutrino Telescopes}

Detecting an absorption feature in the high-energy cosmic neutrino spectrum will be challenging for a number of reasons. Most importantly, a very large number of neutrinos would need to be detected before the spectrum could be adequately reconstructed. In the TeV energy region, in which we have focused on in this letter, next generation kilometer-scale neutrino telescopes, such as IceCube, will detect $\sim$10 muon events per energy decade (from charged-current muon neutrino interactions) per year if an optimistic neutrino flux is assumed (saturating the Waxman-Bahcall bound \cite{wb}, for example) \cite{review}. Of these, only those events which can be identified (by timing and directional considerations) to have originated from high-redshift sources will be of use in detecting an absorption feature. Furthermore, each muon track reveals, at most, the energy of the muon produced, and not the total energy of the responsible neutrino. To measure the cosmic neutrino spectrum with sufficient resolution to detect an absorption feature would require a detector capable of measuring the energy of both the muon track and the accompanying hadronic shower. Alternatively, charged-current electron or tau neutrino interactions could be used, which produce showers containing all of the neutrino's energy. Such events could, however, be confused with showers produced through neutral current interactions.

Assuming good neutrino energy resolution (using well reconstructed, contained events), it may be possible to reliably detect (or exclude) the presence of an absorption line. Even with a low threshold, kilometer-scale experiment and an optimistic neutrino flux, such a determination would likely require many years of exposure, however. To convincingly resolve such a feature would require an experimental program with a greater effective volume and energy resolution than current or planned experiments. 

\begin{figure}

\resizebox{8.5cm}{!}{\includegraphics{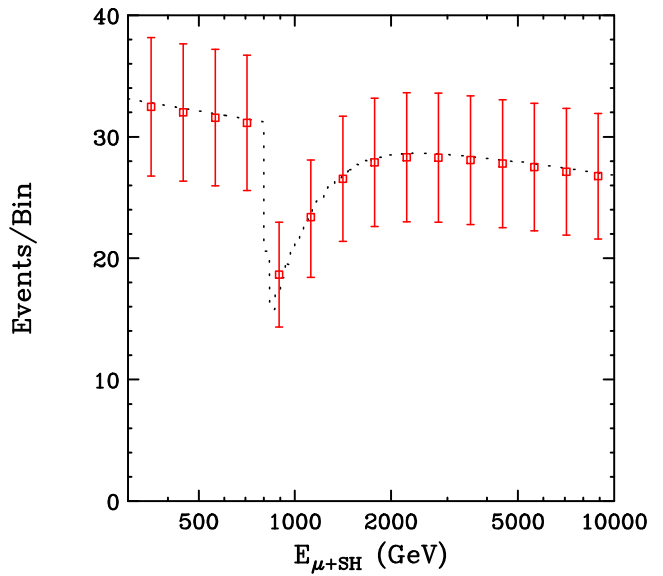}}
\caption{The distribution of events observed in a high-resolution ($\Delta(\log E) \approx 0.1$), high-energy neutrino telescope for the case of $m_U=1$ MeV, $g_{U\nu\nu}=4 \times 10^{-6}$ and sources at redshift $z \approx 5$. We have considered a flux of muon neutrinos of $E_{\nu_{\mu}}^2 dN_{\nu_{\mu}}/dE_{\nu_{\mu}} = 2 \times 10^{-7}$ GeV cm$^{-2}$ s$^{-1}$, a cubic kilometer detector volume and ten years of exposure time.}
\label{bins}
\end{figure}

In Fig.~\ref{bins}, we illustrate what would be required of a future high-energy neutrino telescope to observe an absorption feature by plotting the number of events averaged in $\Delta (\log E) \approx 0.1$ energy bins at energies around an absorption feature. Here, we have used $m_U=1$ MeV, $g_{U\nu\nu}=4 \times 10^{-6}$ and have considered sources at a redshift of $z \approx 5$. We have used an optimistic high-redshift cosmic neutrino flux of $E_{\nu_{\mu}}^2 dN_{\nu_{\mu}}/dE_{\nu_{\mu}} = 2 \times 10^{-7}$ GeV cm$^{-2}$ s$^{-1}$ and an experiment with a cubic kilometer instrumented volume, and ten years of observation time.

\section{Conclusions}

In this letter, we have pointed out that the existence of a new MeV-scale gauge boson coupled to neutrinos would lead to absorption features in the high-energy cosmic neutrino spectrum. The very narrow resonance at a center-of-mass energy equal to the gauge boson's mass will deplete the neutrino flux over an energy range given by $E_R/(1+z) < E_{\nu} < E_R$, where $E_R=m^2_U/2 m_{\nu}$ and $z$ is the redshift of the source. The depletion of the flux in this range will be efficient if the gauge boson's coupling to neutrinos is larger than approximately $g_{U \nu \nu} \gsim 10^{-5} (m_U/\rm{MeV})$.

The possible existence of a light gauge boson has previously been motivated by the observation of 511 keV photons from the galactic bulge, which could be generated by MeV-scale dark matter annihilating through the exchange of the new gauge boson. If such scenario is indeed found in nature, high-energy neutrino astronomy may be able to offer complementary insights into the nature of dark matter.

\bigskip

This work has been supported by the US Department of Energy and by NASA grant NAG5-10842.


\begin{thebibliography}{}


\bibitem{integral}
  P.~Jean {\it et al.},
  Astron.\ Astrophys.\  {\bf 407}, L55 (2003)
  [arXiv:astro-ph/0309484].

\bibitem{511dark}
 C.~Boehm, D.~Hooper, J.~Silk, M.~Casse and J.~Paul,
  Phys.\ Rev.\ Lett.\  {\bf 92}, 101301 (2004)
  [arXiv:astro-ph/0309686].

\bibitem{beacom}
  J.~F.~Beacom and H.~Yuksel,
  Phys.\ Rev.\ Lett.\  {\bf 97}, 071102 (2006)
  [arXiv:astro-ph/0512411];
  J.~F.~Beacom, N.~F.~Bell and G.~Bertone,
  Phys.\ Rev.\ Lett.\  {\bf 94}, 171301 (2005)
  [arXiv:astro-ph/0409403].


\bibitem{lee}
  B.~W.~Lee and S.~Weinberg,
  Phys.\ Rev.\ Lett.\  {\bf 39}, 165 (1977).

\bibitem{lightok}
  C.~Boehm, T.~A.~Ensslin and J.~Silk,
  J.\ Phys.\ G {\bf 30}, 279 (2004)
  [arXiv:astro-ph/0208458].



\bibitem{susycase}
  D.~Hooper and T.~Plehn,
  Phys.\ Lett.\ B {\bf 562}, 18 (2003)
  [arXiv:hep-ph/0212226];
  A.~Bottino, F.~Donato, N.~Fornengo and S.~Scopel,
  Phys.\ Rev.\ D {\bf 68}, 043506 (2003)
  [arXiv:hep-ph/0304080].

\bibitem{nmssm}
  J.~F.~Gunion, D.~Hooper and B.~McElrath,
  Phys.\ Rev.\ D {\bf 73}, 015011 (2006)
  [arXiv:hep-ph/0509024].

\bibitem{scalar}
  C.~Boehm and P.~Fayet,
  Nucl.\ Phys.\ B {\bf 683}, 219 (2004)
  [arXiv:hep-ph/0305261].


\bibitem{fayet}
  P.~Fayet,
  Phys.\ Rev.\ D {\bf 70}, 023514 (2004)
  [arXiv:hep-ph/0403226].


\bibitem{collider}
  N.~Borodatchenkova, D.~Choudhury and M.~Drees,
  Phys.\ Rev.\ Lett.\  {\bf 96}, 141802 (2006)
  [arXiv:hep-ph/0510147];
  B.~McElrath,
  Phys.\ Rev.\ D {\bf 72}, 103508 (2005)
  [arXiv:hep-ph/0506151];
  P.~Fayet,
  arXiv:hep-ph/0607094;
  M.~Ablikim {\it et al.}  [BES Collaboration],
  Phys.\ Rev.\ Lett.\  {\bf 97}, 202002 (2006)
  [arXiv:hep-ex/0607006];
  P.~Fayet,
  Phys.\ Rev.\ D {\bf 74}, 054034 (2006)
  [arXiv:hep-ph/0607318].


\bibitem{boehmneutrino}
  C.~Boehm,
  Phys.\ Rev.\ D {\bf 70}, 055007 (2004)
  [arXiv:hep-ph/0405240].



\bibitem{atomic}
  C.~Bouchiat and P.~Fayet,
  Phys.\ Lett.\ B {\bf 608}, 87 (2005)
  [arXiv:hep-ph/0410260].


\bibitem{sn}
  P.~Fayet, D.~Hooper and G.~Sigl,
  Phys.\ Rev.\ Lett.\  {\bf 96}, 211302 (2006)
  [arXiv:hep-ph/0602169].



\bibitem{zburst}
  T.~J.~Weiler,
  Astropart.\ Phys.\  {\bf 11}, 303 (1999)
  [arXiv:hep-ph/9710431];
  B.~Eberle, A.~Ringwald, L.~Song and T.~J.~Weiler,
  Phys.\ Rev.\ D {\bf 70}, 023007 (2004)
  [arXiv:hep-ph/0401203].

\bibitem{goldberg}
  H.~Goldberg, G.~Perez and I.~Sarcevic,
  JHEP {\bf 0611}, 023 (2006)
  [arXiv:hep-ph/0505221].


\bibitem{wb}
  E.~Waxman and J.~N.~Bahcall,
  Phys.\ Rev.\ D {\bf 59}, 023002 (1999)
  [arXiv:hep-ph/9807282].

\bibitem{review}
For a review of high-energy neutrino astronomy, see:
  F.~Halzen and D.~Hooper,
  Rept.\ Prog.\ Phys.\  {\bf 65}, 1025 (2002)
  [arXiv:astro-ph/0204527].









\end{thebibliography}
\end{document}